\documentstyle[amssymb,aps,epsf]{revtex}
\begin{document}
\title{On high $Q^{2}$ behavior of the pion form factor for transitions
$\gamma^{\ast} \gamma \to \pi^{0}$ and $\gamma ^{\ast} \gamma^{\ast} \to
\pi^{0}$ within the nonlocal quark-pion model.}
\author{I.V. Anikin$^{1}$, A. E. Dorokhov$^{1,2}$, Lauro Tomio$^2$}
\address{$^{1}$ Bogoliubov Laboratory for Theoretical Physics, Joint\\
Institute for Nuclear Research, 141980, Dubna, Russia \\
$^{2}$ Instituto de F\'{\i}sica Te\'orica, UNESP,\\
Rua Pamplona, 145, 01405-900, S\~ao Paulo, Brazil}
\date{\today}
\maketitle

\begin{abstract}
The behavior of the transition pion form factor for processes $\gamma
^{\star}\gamma \rightarrow \pi ^{0}$ and $\gamma ^{\star }\gamma
^{\star}\rightarrow \pi ^{0}$ at large values of space-like photon momenta
is estimated within the nonlocal covariant quark-pion model. It is shown
that, in general, the coefficient of the leading asymptotic term depends
dynamically on the ratio of the constituent quark mass and the average
virtuality of quarks in the vacuum and kinematically on the ratio of photon
virtualities. The kinematic dependence of the transition form factor
allows us to obtain the relation between the pion light-cone distribution
amplitude and the quark-pion vertex function. The dynamic dependence
indicates that the transition form factor $\gamma ^{\star}\gamma \rightarrow
\pi ^{0}$ at high momentum transfers is very sensitive to the
nonlocality size 
of nonperturbative fluctuations in the QCD vacuum. \newline
%PACS number(s): {12.38.Aw, 12.38.Lg, 14.40.Aq, 11.10.Hi, 11.10.Lm}
%11.10.Hi Renormalization group evolution of parameters
%11.10.Lm Nonlinear or nonlocal theories and models
%12.38.Aw General properties of QCD (dynamics, confinement, etc.)
%12.38.Lg Other nonperturbative calculations
%14.40.Aq pi, K, and eta mesons
\end{abstract}

\section{Introduction}

The interest in the form factor $T_{\pi ^{0}}(q_{1}^{2},q_{2}^{2})$ for
transition processes $\gamma ^{\star }(q_{1})\gamma (q_{2})\rightarrow \pi
^{0}(p)$ and $\gamma ^{\star }(q_{1})\gamma ^{\star }(q_{2})\rightarrow \pi
^{0}(p)$, where $q_{1}$ and $q_{2}$ are photon momenta, has again increased
recently. Experimentally, the data on the form factor $T_{\pi ^{0}}$ for
small virtuality of one of the photons, $q_{2}^{2}\approx 0$, with the
virtuality of the other photon being scanned up to $8$ GeV$^{2}$, are known
from CELLO \cite{CELLO} and CLEO \cite{CLEO} Collaborations. Theoretically,
at zero virtualities, the form factor $T_{\pi ^{0}}\left( 0,0\right) $ is
related to the axial anomaly. At asymptotically large photon virtualities,
the behavior is predicted by perturbative QCD
(pQCD)~\cite{BrLep79}-\cite{Braaten}
(see \cite{RadMus97} for recent discussions) and depends crucially on the
internal pion dynamics that is parametrized by the nonperturbative pion
distribution amplitude (DA), $\varphi^{A}_\pi(x)$,
with $x$ being the fraction of the pion momentum, $p$, carried by a quark.
Moreover, the knowledge of the
off-shell structure of the form factor enables one to significantly reduce
the uncertainty in the evaluation of the hadronic light-by-light scattering
contribution to the muon $g-2$~\cite{Kinoshita98}, which is relevant for the
current experiment E821 at BNL \cite{E821}.

In the following, it is convenient to parametrize photon virtualities as
 $q_{1}^{2}=-(1+\omega )Q^{2}/2$, $q_{2}^{2}=-(1-\omega )Q^{2}/2$, where
$Q^2$
and $\omega$ are, respectively, the total virtuality of the photons and the
asymmetry in their distribution:
\begin{equation}
Q^{2}=-(q_{1}^{2}+q_{2}^{2})\geq 0, \;\;\;\;{\rm and}\;\;\;\; \omega
=(q_{1}^{2}-q_{2}^{2})/(q_{1}^{2}+q_{2}^{2}).  \label{Omega}
\end{equation}
The experimental data from CLEO \cite{CLEO} for the process $\gamma
^{\ast}\gamma \rightarrow \pi ^{0}$ $\left( |\omega |=1\right) $ can be
fitted by a monopole form factor:
\begin{equation}
\left. T_{\pi ^{0}}(q_{1}^{2}=-Q^{2},q_{2}^{2}=0)\right| _{fit}=
\frac{g_{\pi\gamma \gamma }}{1+Q^{2}/\Lambda _{\pi }^{2}}, \ \ \ \ \ \Lambda
_{\pi}\simeq 0.77\ GeV,  \label{Fpiggfit}
\end{equation}
where $g_{\pi \gamma \gamma }$ is the two-photon pion decay constant. In the
 lowest order of pQCD, by using the light-cone Operator Product
Expansion (OPE),
 the high $Q^{2}$ behavior of the form factor is
predicted~\cite{BrLep79,Chase} as
\begin{equation}
\left. T_{\pi ^{0}}(q_{1}^{2},q_{2}^{2})\right| _{Q^{2}\rightarrow \infty}
=J\left( \omega \right) \frac{f_{\pi }}{Q^{2}} +O(\frac{\alpha_s}{\pi})+
O(\frac{1}{Q^4}),  \label{AmplAsympt}
\end{equation}
with the asymptotic coefficient given by
\begin{equation}
J\left( \omega \right) =\frac{4}{3}
\int_{0}^{1} \frac{dx}{1-\omega^2 ( 2x-1)^2}\varphi_\pi^{A}(x),  \label{J}
\end{equation}
where $f_{\pi }=93$ {\rm MeV} is the weak pion decay constant and
the leading-twist pion light-cone DA is normalized by
$\int_{0}^{1} dx\varphi_\pi^{A}(x)=1$.
Since the pion DA reflects the internal nonperturbative pion dynamics,
the prediction of the value of  $J\left( \omega \right)$
is rather a nontrivial task,
and its accurate measurement would provide quite valuable information.

 It is important to note that for the transition process considered, the
leading asymptotic term of pQCD expansion (\ref{AmplAsympt}) is not
suppressed by the strong coupling constant $\alpha_s$. Hence, the pQCD
prediction (\ref{AmplAsympt})
may become reasonable\footnote{
 This is in contrast with the case of electromagnetic form factors and
the wide-angle Compton scattering process ({\it e.g.}, see
Ref.~\cite{ILSR}), where the soft overlap
contributions are important at moderately high $Q^{2}$.}
at the highest of the presently accessible momenta $Q^{2}\sim 10$ GeV$^2$.
 At asymptotically high $Q^{2}$, the DA evolves to
$\varphi_\pi^{A,asympt}(x)$
$=6x(1-x)$ and $J_{asympt}\left( \left| \omega \right| =1\right) =2$.
The fit of CLEO data ~(\ref{Fpiggfit}) corresponds to $J_{CLEO}\left( \left|
 \omega \right| \approx 1\right) \simeq 1.6$ indicating that already at
moderately high momenta this value is not too far from its asymptotic
limit.

 However, since the pQCD evolution of the DA reaches the asymptotic
regime very
slowly, its exact form at moderately high $Q^2$ may not coincide with
$\varphi_\pi^{A,asympt}(x)$.
At lower $Q^{2}$, the power corrections to the form factor become important.
Thus, to study the behavior of the transition form factor, at experimentally
accessible $Q^{2}$, is the subject of nonperturbative dynamics, where
the same type of the leading high $Q^{2}$ behavior, as in eq. (\ref
{AmplAsympt}), was obtained by different methods. So, the theoretical
determination of the transition form factor is still challenging, and it is
desirable to obtain direct calculations of $T_{\pi
^{0}}(q_{1}^{2},q_{2}^{2}) $, without any {\it a priori} assumptions about
the shape of the pion DA.

 The transition form factor in the symmetric kinematics,
$q_{1}^{2}=q_{2}^{2}$, at high  virtualities was considered in
\cite{NSVVZ84} by using the local OPE with the result
 $J_{OPE}\left( \omega=0\right) =4/3$ for the asymptotic coefficient,
which is in agreement with prediction from (\ref{J}) at $\omega=0$.
Within the local OPE, one can represent
 $J_{OPE}(\omega)$ as an expansion in powers of $\omega^2$, with the
 coefficients of expansion given by the moments of the pion DA:
$\int^1_0 dx (2x-1)^{2n} \varphi_\pi^{A}(x)$. In \cite{manohar90} (see
also \cite{gorsky}), it was shown that the
local OPE is well convergent in the kinematic region, when the photon
virtualities are close to each other: $\left| \omega \right| \lesssim 1/2$.
 In this kinematics, the result for the asymptotic coefficient is still
close to 4/3. However, in these references it was pointed out that for
$|\omega |\gtrsim1/2$  potentially large corrections to the first term
of the local OPE \cite{manohar90,gorsky}
 and also to the light-cone pQCD \cite{BrLep79} predictions exist at any
finite $Q^{2}$.  With increasing $|\omega|$, the number of terms of OPE
with higher-dimension vacuum expectation
 values growths rapidly, but it is practically a hopeless task to find
more than few terms of the local expansion. Therefore, much more
detailed information about the nonperturbative QCD
vacuum is necessary to have control over the operator expansion.

In ref.~\cite{MikhRad90}, some progress was achieved, by using a refined
 technique based on the OPE with nonlocal condensates~ \cite{MihRad92},
which is equivalent to
inclusion of the whole series of power corrections. By using the QCD sum
 rules with nonlocal condensates, it was shown that this approach works
in almost the whole kinematic
region $\left| \omega \right| \lesssim 1$, and that for high values of the
asymmetry parameter $\left| \omega \right| \gtrsim 0.8$, the pion transition
form factor is very sensitive to the nonlocal structure of the QCD vacuum.
 The latter is characterized by the average quark virtuality in the
vacuum \cite{MihRad92},
$\lambda _{q}^{2}$,  and, within the instanton model \cite{Shuryak96},
may be expressed through the average instanton size, $\rho _{c}$, as $\lambda
_{q}^{2}\approx 2\rho _{c}^{-2}$ \cite{DEM97,LauroDo98}.
 In \cite{RadRus96}, the  form factor $\gamma ^{\ast }\gamma \rightarrow
\pi ^{0}$ was
directly calculated from a QCD sum rule for the three-point function, leading
to the estimate $J_{QCDsr}\left( \omega =1\right) \approx 1.6\pm 0.3$.

 The covariant nonlocal low-energy models (see, {\it e.g.},
\cite{Holdom89,Gross90}),
based on the Schwinger-Dyson (SD) approach to dynamics of quarks and gluons,
have many attractive features, as
the approach is consistent with the low-energy theorems. In particular, the
Abelian axial anomaly is within this approach, and the standard result for
$T_{\pi ^{0}}(0,0)\equiv g_{\pi ^{0}\gamma \gamma }=\left(4\pi ^{2}f_{\pi
}\right) ^{-1}$ is reproduced exactly.  Within this nonperturbative model
of quark-pion interaction,
both the small mass and composite structure of the pion are realistically
described. Furthermore, the intrinsic nonlocal structure of the model may be
motivated by fundamental QCD processes like the instanton and gluon exchanges.
In \cite{Gross90} the transition form factor
$\gamma ^{\ast }\gamma \rightarrow \pi ^{0}$ was considered at low $Q^2$ and
 agreement with data was obtained. There, it was observed that the
results are very sensitive to the value of constituent quark mass.

 In this letter, within the covariant nonlocal low-energy model of the
quark-pion interaction
we study the high $Q^2$ behavior of the pion transition form factor
$\gamma ^{\ast }\gamma^{\ast } \rightarrow \pi ^{0}$
 in general kinematics. We show that the asymptotic coefficient $J\left(
\omega \right)$, as demonstrated in QCD sum rules
\cite{MikhRad90,RadRus96}, depends on the kinematics of the
 transition process and on the internal pion dynamics induced by the
nonlocal structure
of the QCD vacuum. The
dynamic dependence of $J$ is governed by the so-called diluteness
parameter $M_{q}/\lambda _{q}$, where $M_{q}$ is the constituent quark mass.
 When considering the model dependence of the asymptotic coefficient
$J$, experimental data can be very useful to distinguish between
different assumptions made on nonperturbative
dynamics of the QCD vacuum. Within the nonlocal quark-pion model
the expression for the asymptotic coefficient $J$ is found
in the whole kinematic region of $\omega$. Moreover, from this dependence,
the pion DA is reconstructed in terms of the quark-pion vertex function.

\section{Effective quark-pion model and pion transition $\protect\gamma
^{\ast }\protect\gamma ^{\ast }\rightarrow \protect\pi ^{0}$ form factor}

 The effective quark-pion dynamics can be summarized in the covariant
nonlocal action given by
\begin{equation}
S_{int}=-\int d^{4}xd^{4}y\ F\left[x+y/2,x-y/2;\Lambda ^{-2}\right]
 \bar{q}(x+y/2)\ [M_{q}+g_{\pi \bar{q}q}i\gamma _{5}\tau ^{a}\pi
^{a}(x)]\ q(x-y/2),
\label{S_Q_Pi}
\end{equation}
where the dynamic vertex $F\left[x+y/2,x-y/2;\Lambda ^{-2}\right]$
 with nonlocality size $\Lambda^{-1}$ depends on the coordinates of the
quark and antiquark; $q(x)$ and $\pi(x)$ are, respectively, the quark
and pion fields. The nonlocal
vertex characterizes the coordinate dependence of order parameter for
spontaneous chiral-symmetry breaking and can be expressed in terms of the
nonlocal quark condensates.

In the following calculations, we restrict ourselves to the approximation
(see, e.g.  \cite{Gross90})
\begin{equation}
F\left[x+y/2,x-y/2;\Lambda ^{-2}\right] \to F(y^2, \Lambda^{-2}),
\label{Separable}\end{equation}
when the dynamic quark-pion vertex depends only
 on the relative coordinate of the quark and antiquark squared, $y^2$,
if neglecting
the dependence of the vertex  on angular variable $(yx)$.
The Fourier transform of the vertex function in
the Minkowski space is defined as $\tilde F(k^2;\Lambda^{2}) = \int d^4x
F(x^{2};\Lambda ^{-2})\exp(-ikx)$ with normalization $\tilde F
(0;\Lambda^{2}) = 1$, and we assume that it rapidly decreases in the
Euclidean region ($k^2=-k_E^2\equiv -u$). We also
approximate the momentum-dependent quark self-energy in the
 quark propagator $S^{-1}(k)=\widehat{k}-M_{q}$ by a constant
mass~\cite{Gross90} and neglect small effects of the pion mass.
We have to note that the approximations used are not fully consistent. In
particular, due to neglecting the momentum dependence of the quark mass,
some low-energy theorems are violated. Further, as we show below, the choice
of the model for the quark-pion vertex (\ref{Separable}) depending
 only on the relative coordinate induces a certain artifact in the $x$
behavior of DA (see below).
However, these deficiencies of the approximation chosen are not essential for
the present purposes and do not lead to essential numeric errors.

The quark-pion coupling is given by the compositeness condition~\cite{Gross90}
\begin{equation}
g_{\pi \bar{q}q}^{-2}=\frac{N_{c}}{8\pi ^{2}}\int_{0}^{\infty }duu\tilde F
^{2}(-u;\chi ^{-2})\frac{3+2u}{\left( 1+u\right) ^{3}};  \label{gpiqq}
\end{equation}
and the pion weak decay constant is expressed by
\begin{equation}
f_{\pi }=\frac{N_{c}g_{\pi \bar{q}q}}{4\pi ^{2}}M_{q}\int_{0}^{\infty} duu
\tilde F(-u;\chi ^{-2})\frac{1}{\left( 1+u\right) ^{2}}.  \label{f_pi}
\end{equation}
We have rescaled the integration variable by the quark mass squared and
introduced the parameter $\chi=M_{q}/\Lambda $ that characterizes the
diluteness of the QCD vacuum. Within the instanton vacuum model, the size of
nonlocality of the nonperturbative gluon field, $\rho _{c}\sim \Lambda
^{-1}$, is much smaller than the quark Compton length $M_{q}^{-1}$; thus,
$\chi$ is a small parameter~\cite{Shuryak96}.

Let us consider the contribution to
the $\gamma ^{\ast }\gamma ^{\ast }\pi ^{0}$ invariant amplitude as
calculated from the triangle diagrams:
\[
M\left( \gamma ^{\ast }\left( q_{1},e_{1}\right) \gamma ^{\ast} \left(
q_{2},e_{2}\right) \rightarrow \pi ^{0}\left( p\right) \right) =m_{\pi
\gamma \gamma }(q_{1},e_{1};q_{2},e_{2})+ m_{\pi \gamma \gamma
}(q_{2},e_{2};q_{1},e_{1})
\]
where $e_{i}(i=1,2)$ are the photon polarization vectors, and
\begin{equation}
m_{\pi \gamma \gamma }(q_{1},e_{1};q_{2},e_{2})=-\frac{N_{c}}{3} g_{\pi
qq}\int \frac{d^{4}k}{(2\pi )^{4}} \tilde F(k^{2};\Lambda ^{2})tr\{i\gamma
_{5} S(k-p/2){\hat{e}}_{2}S[k-(q_{1}-q_{2})/2]{\hat{ e}}_{1}S(k+p/2)\}.
\end{equation}
If the tensor $\epsilon _{\mu \nu \rho \sigma}e_{1}^{\mu}e_{2}^{\nu }
q_{1}^{\rho }q_{2}^{\sigma }$ is factorized from this amplitude,
the form factor can be expressed as
\begin{equation}
T_{\pi ^{0}}(q_{1}^{2},q_{2}^{2})=\frac{g_{\pi qq}}{2\pi ^{2}}M_{q}I_{\pi
\gamma \gamma }(q_{1}^{2},q_{2}^{2},p^{2}),  \label{invAmpl}
\end{equation}
where the Feynman integral $I_{\pi \gamma \gamma}
(q_{1}^{2},q_{2}^{2},p^{2}) $ is given by
\begin{equation}
I_{\pi \gamma \gamma }(q_{1}^{2},q_{2}^{2},p^{2})=\int \frac{d^{4}k}{i\pi^{2}}
\frac{\tilde F(k^{2};\Lambda ^{2})}{[M_{q}^{2}-(k+p/2)^{2}-i\varepsilon]
[M_{q}^{2}-(k-p/2)^{2}-i\varepsilon]
[M_{q}^{2}-(k-(q_{1}-q_{2})/2)^{2}-i\varepsilon ]}.  \label{Ipigg}
\end{equation}
Within the constant-mass approximation, the low-energy relation
following from the Adler - Bell - Jackiw (ABJ) axial anomaly [$f_{\pi }g_{\pi
\gamma \gamma }={1}/{(4\pi ^{2})}$] is well reproduced numerically with a
relative error smaller than $10\%$ ~\cite{Gross90}. In the formal limit of a
very dilute vacuum medium $\chi <<1$, the results are consistent with the
ABJ anomaly and the Goldberger-Treiman relation.

 Let us note that the integral (\ref{Ipigg}) is similar in structure to
the integral arising in the lowest order of pQCD treating the
quark-photon interaction perturbatively.
In the latter case, its asymptotic behavior is due to the subprocess
$\gamma^*(q_1) + \gamma^*(q_2) \to \bar q(\bar xp) + q (xp) $
with $x$ ($\bar x$) being  the fraction of the pion momentum $p$ carried
by the quark produced at the $q_1$ ($q_2)$ photon vertex.
 The relevant diagram is similar to the handbag diagram for hard
exclusive processes, with the main difference that one should use, as a
nonperturbative input, the
quark-pion vertex instead of the pion DA. As we see below, this similarity
allows one to translate the form of the quark-pion vertex into a specific
shape of the pion DA.

\section{Pion transition $\protect\gamma ^{\ast }\protect\gamma ^{\ast}
\rightarrow \protect\pi ^{0}$ form factor at moderately high $Q^{2}$}

In this section, we estimate the asymptotics of the transition
form factor. To this end, we rewrite the expression for integral (\ref{Ipigg})
in the form that is obtained after rotating to the Euclidean space
[$k^{2}\rightarrow -u,$ $-id^{4}k\rightarrow \pi ^{2}udu, $ $\tilde F
(k^{2};\Lambda ^{2})\rightarrow \tilde F(-u;\Lambda ^{2})$], by using the
 Feynman $\alpha -$parameterization for the denominators and integrating
over the angular variables. Then, the corresponding integral $I_{\pi
\gamma \gamma }$
is given by
\begin{equation}
I_{\pi \gamma \gamma }(q_{1}^{2},q_{2}^{2},p^{2})=\int_{0}^{\infty }\frac{
duu\tilde F(-u;\Lambda ^{2})}{M_{q}^{2}+u-\frac{p^{2}}{4}}
\int_{0}^{1}d\alpha \left[ \frac{1}{\sqrt{b^{4}-a_{+}^{4}}\left( b^{2}+\sqrt{
b^{4}-a_{+}^{4}} \right) }+\frac{1}{\sqrt{b^{4}-a_{-}^{4}}\left( b^{2}+\sqrt{
b^{4}-a_{-}^{4}} \right) }\right] ,  \label{Igpigg3}
\end{equation}
where
\begin{equation}
b^{2}=M_{q}^{2}+u+\frac{1}{2}\alpha Q^{2}-\frac{1}{4}\left( 1-2\alpha
\right) p^{2},\quad a_{\pm }^{4}=2u\alpha Q^{2}\left( \alpha \pm \omega
\left( 1-\alpha \right) \right) -\left( 1-2\alpha \right) up^{2}.
\label{notas2}
\end{equation}
In this way, the expression (\ref{Igpigg3}) can be safely analyzed in the
 asymptotic limit of high total virtuality of the photons
$Q^{2}\rightarrow \infty $. Moreover, the integral over $\alpha $ can be
taken analytically, leading, in
the chiral limit $m_{\pi }=0$, to the asymptotic expression given by
eq.~(\ref{AmplAsympt}), where
\begin{eqnarray}
&&J(\omega)\equiv J_{np}\left( \omega \right) =\frac{2N_\pi}{3\omega }
\left\{ {\ \int_{0}^{\infty}du\ \frac{\tilde F(-u;\chi ^{-2})}{1+u} \ln
\left[ \frac{1+u\left( 1+\omega \right)}{1+u\left( 1-\omega \right) } \right]
}\right\}  \label{Igpigg2} \\
&&N_\pi = \left[ {\ \int_{0}^{\infty}duu \frac{\tilde F(-u;\chi ^{-2})}
{\left( 1+u\right) ^{2}}}\right] ^{-1}.  \nonumber
\end{eqnarray}
The integrand in the numerator of (\ref{Igpigg2}) is quite different from the
integrand defining the decay constant $f_{\pi }$, given in eq.~(\ref{f_pi}).
 From eq. (\ref{Igpigg2}) it is clear, that the prediction of the
nonperturbative
approach about the asymptotic coefficient is rather sensitive to the product
$\chi$ of the value of the constituent mass $M_{q}$ and the size of
nonlocality  $\Lambda^{-1}$ of the vertex $F(x^2;\Lambda^{-2})$ and to
the relative distribution of the total virtuality among photons, $\omega $.
In particular, for the off-shell process $\gamma ^{\ast }\gamma ^{\ast
}\rightarrow \pi ^{0}$ in the kinematic case of symmetric distribution of
photon virtualities, $q_{1}^{2}=q_{2}^{2}\rightarrow -\infty $ ($\omega
\rightarrow 0$), the result obtained from eq.~(\ref{Igpigg2}) is
$J\left( \left| \omega \right| =0\right) =4/3$ in agreement
with the OPE prediction~\cite{NSVVZ84}.

Let us note, that we use an approximation to the model with constant
constituent quark masses for all three quark lines in the diagram of the
process. However, the asymptotic result (\ref{Igpigg2}) is independent of
the value of the mass parameter in the quark propagator with hard momentum
flow, as it should be. The other two quark lines remain soft during the
 process; thus, the mass parameter $M_{q}$ may be considered as given on
a certain  characteristic soft scale in the momentum-dependent case
$M_{q}\left( \lambda _{q}^{2}\right) $. It means that the dynamic and
kinematic dependence of $J_{np}\left( \omega \right) $ found in
(\ref{Igpigg2}) will
be unchanged, even if one includes the momentum dependence of the quark mass
and considers the dressed quark-photon vertex which goes into the bare one,
$\gamma ^{\mu }$, as one of the squared quark momenta becomes infinite.

 Both the expressions for $J$ derived within the nonlocal quark-pion
model (\ref{Igpigg2})
 and from the light-cone OPE (\ref{J}) can be put into the common form
\begin{equation}
J\left( \omega \right) =\frac{2}{3\omega }\int_{0}^{1}\ d\xi R(\xi )\ln
\left[ \frac{1+\xi \omega }{1-\xi \omega }\right]   \label{R}
\end{equation}
with
\begin{equation}
R_{pQCD}(\xi )=-\frac{d}{d\xi }\varphi _{\pi }^{A}\left( \frac{1+\xi }{2}
\right) \;\;\;\;{\rm and}\;\;\;\;R_{np}(\xi )=N_{\pi }\tilde{F}\left( \frac{
-\xi }{1-\xi };\chi ^{-2}\right) \frac{1}{1-\xi },\;\;\;{\rm where}
\;\;\;0\leq \xi \equiv (2x-1)\leq 1  \label{WF_VFdif}
\end{equation}
and similar expressions for $-1\leq \xi \leq 0$. Equating both contributions,
 we find the pion DA in terms of the vertex function on a certain
low-energy scale
$\mu _{0}^{2}\sim \Lambda ^{2}$
\begin{equation}
\varphi _{\pi }^{A}(x)=N_{\pi }\int_{|2x-1|}^{1}\frac{dy}{1-y}\tilde{F}
\left( \frac{-y}{1-y};\chi ^{-2}\right) .  \label{WF_VF}
\end{equation}
 Thus, we show that eq.  (\ref{Igpigg2}) obtained within the nonlocal
quark-pion model is equivalent to the standard lowest-order pQCD result
(\ref{J}), with the
only difference that the nonperturbative information
accumulated in the pion DA $\varphi _{\pi }^{A}(x)$ is  represented
by the quark-pion vertex function $\tilde F(-u;\chi^{-2})$.

 We have to note that an explicit form of the asymptotic coefficient
(\ref{Igpigg2}) and the relation between the DA and the vertex function
 depend on the model of quark-pion interaction (\ref{S_Q_Pi}). In
particular, the expression (\ref{WF_VF}) is obtained within the
approximation (\ref{Separable}), when the quark-pion vertex depends only
on the relative coordinate. This approximation results
in the artificial dependence of DA on the modulo function of  $x$ and leads
to the nonsmooth behavior of the distribution at $x=1/2$ (see Fig. 1).
 These peculiarities disappear if the angular dependence of the vertex
motivated by, {\it e.g.}, the instanton model is recovered
\footnote{
In \cite{LauroDo98}, emerging of a similar cusp for the pion distribution
function and its disappearance, if the angular dependence in the
vertex is taken into account, were demonstrated.}.

Let us estimate a realistic value for the diluteness parameter $\chi $ and
check if the model under consideration is consistent with CLEO data. The
vertex function $\tilde F(k^{2};\Lambda ^{2})$ phenomenologically describes
the nonlocal structure of the nonperturbative QCD vacuum and may be modeled
within the instanton vacuum model \cite{LauroDo98}. For the present purpose,
the vertex function can be well approximated by the Gaussian form $\tilde F
(k^{2};\Lambda ^{2})=\exp (k^{2}/\Lambda ^{2})$. The inverse size of the
vertex nonlocality, $\Lambda$, is naturally related to the average
virtuality of quarks that flow through the vacuum, $\lambda _{q}^{2}$,
\cite{MihRad92,DEM97,LauroDo98}
\begin{equation}
\lambda _{q}^{2}\equiv \frac{\langle :\bar{q} D^{2}q:\rangle }{\langle :\bar{
q}q:\rangle }= -\frac{N_{c}M_{q}^{5}}{4\pi ^{2}\langle \bar{q}q\rangle }
\int_{0}^{\infty }duu^{2}\frac{\tilde F(-u;\chi ^{-2})}{u+1},  \label{QVirt}
\end{equation}
where $D_\mu$ is the covariant derivative with respect to the strong gauge
field, and the quark condensate is expressed as
\begin{equation}
\langle \bar{q}q\rangle =-\frac{N_{c}M_{q}^{3}}{4\pi ^{2}}
\int_{0}^{\infty}duu\frac{\tilde F(-u;\chi ^{-2})}{u+1}.  \label{QQcond}
\end{equation}
The value of $\lambda _{q}^{2}$ is known from the QCD sum rule analysis,
$\lambda _{q}^{2}$ $\approx 0.5\pm 0.1{\rm GeV}^{2}$ \cite{BI82}. For the
Gaussian vertex, one has $\Lambda ^{2}\approx \lambda _{q}^{2}$ if
$\Lambda^2/M_q^2 >1$. The quark mass parameter $M_{q}$ is given by the
Goldberger-Treiman relation $M_{q}=g_{\pi qq}f_{\pi }$, with the quark-pion
coupling being fixed by the compositeness condition (\ref{gpiqq}), and its
value is consistent with $M_{q}=250\sim 300$ {\rm MeV}. Varying the model
parameters within the intervals $\Lambda ^{2}=0.55\pm 0.05$ ${\rm GeV}^{2}$
and {$M_{q}=275\pm 25$ {\rm MeV}}, we have $\lambda _{q}^{2}=0.65\pm 0.05$
${\rm GeV}^{2}$, $\langle \bar{q}q\rangle =-\left( 205\pm 15{\ {\rm MeV}}
 \right) ^{3}$ and $J_{np}\left( \omega =1\right) =1.80\pm 0.05$. When
taking into account the error in the experimental fit, this estimate is
in agreement with the CLEO data. It also agrees with the new estimate
$J_{QCDsr}\left( \omega =1\right) \approx 1.83\pm 0.05$ made in
\cite{MikhBak98} by the QCD sum rules with nonlocal condensates.

It is instructive to consider some extreme cases, depending on the physics
under consideration. If the QCD vacuum were a very dilute vacuum
$M_{q}<<\Lambda $, then the vertex function $\tilde{F}(-u,\chi ^{-2})$ is
 a very slowly decaying function. This corresponds to the local
quark-pion vertex. In that case, the coefficient $J_{dilute}^{np}\left(
\left| \omega \right| =1\right)$ goes logarithmically to infinity and
the DA also becomes flat $\varphi _{\pi}^{A}(x)=1$.
The latter can be seen from Eq. (\ref{WF_VFdif}), where the
 derivative of DA with respect to $x$ is very small, since in this
limit, the normalization factor $N_{\pi }$ becomes small.
From (\ref{WF_VF}) it is clear that the integration region is effectively
restricted from above by the scale $y_{1}=\chi ^{-2}/\left( 1+\chi
^{-2}\right) $ and from below by  $y_{0}=|1-2x|$. These scales are
well separated in the region $x_{0}\lesssim x\lesssim 1-x_{0}$, where 
$x_{0}\approx 1/2\left( 1+\chi ^{-2}\right) $. It means that the DA is
suppressed at the edges of the kinematic interval $1-|1-2x|<2x_{0}$, where
quarks are soft. As it was pointed out earlier, the incorporation of
nonperturbative effects results in the intrinsic transverse structure of
hadronic wave functions \cite{KrJ,AED95}, as well as the Sudakov
perturbative factor \cite{LiSt} modifies the hard scattering picture of
exclusive reactions and essentially improves it. As a result, perturbative
QCD calculations of the hadron form factors extend the kinematic region of
self - consistency from asymptotic values of $Q$ to the region starting from
$Q\sim O(1GeV)$.
In the opposite extreme case of a very dense medium (heavy quark limit,
$M_{q}>>\Lambda $), $J_{dense}^{np}\left( \left| \omega \right| =1\right)
= 4/3$, as it is predicted by the first term in the OPE result
\cite{manohar90}.
 In that case, the limit $y_{1}$ is small, and the integrand in
(\ref{WF_VF}) is
concentrated in the vicinity of $x=1/2$. Thus, the DA  becomes 
 $\varphi_{\pi }^{A}(x)\propto \delta (x-1/2)$, as it is expected. As we
shown above, a realistic situation is in-between these two extremes.

 These different situations are illustrated in terms of the pion DA,
(\ref{WF_VF}), in Fig.1. As it is clear from the figure, the model pion
DA, under the realistic choice of the parameter $\chi \approx 0.4$, is
close to the asymptotic DA. As noticed in the introduction, by
considering the actual accessible data, the nonperturbative dynamics may
dominate. Therefore, the data turn out to be quite restrictive and
uniquely indicate that the dilute
regime is realized in the QCD vacuum. In Fig. 2, for the process $\gamma
\gamma ^{\ast }\rightarrow \pi ^{0}$ ($\omega =1$), we plot the asymptotic
coefficient $J_{np}\left( \omega =1\right) $ as a function of the dynamical
diluteness parameter squared $\chi ^{2}$. In this figure, we indicate the
values of $J_{np}\left( \omega =1\right) $ obtained from CLEO data and model
predictions. In Fig. 3, the asymptotic coefficient $J_{np}\left( \omega
\right) $ is plotted as a function of the kinematic asymmetry parameter
$\omega $, at $\chi ^{2}=0.15$ and $\chi ^{2}=0.35$. The first value of $\chi
$ corresponds to the model estimate; and the second one, to the central point
of the CLEO data fit.

To get further interpretation of eq. (\ref{WF_VF}), we can express the
DA as the transverse momentum integral of the pion light-cone wave function
\cite{BrLep79}
\begin{equation}
\varphi _{\pi }^{A}(x)=\int_{0}^{\infty }d\vec{k}_{\bot }^{2}\Psi _{\pi
}^{A}(x,\vec{k}_{\bot }^{2}).  \label{Psi}
\end{equation}
Rewriting the r.h.s. of (\ref{WF_VF}) via the original variable $u=y/(1-y)$
and then substituting $u$ by the light-cone combination
 $\vec{k}_{\bot}^{2}/(x\bar{x})$, that is the invariant mass of the
$q\bar{q}$ pair
squared, we identify the pion wave function as
\begin{equation}
\Psi _{\pi }^{A}(x,\vec{k}_{\bot }^{2})=\frac{N_{\pi }
\tilde F(-\vec{k}_{\bot}^{2}/(x\bar{x});
\Lambda ^{2})}{x\bar xM_{q}^{2}+\vec{k}_{\bot }^{2}}.
 \Theta\left( \vec{k}_{\bot }^{2}\geq
\frac{|1-2x|x\bar{x}}{1-|1-2x|}M_{q}^{2}\right).
\label{Psi(x,k)}\end{equation}
The vertex function $\tilde F$ in our model of the pion wave function plays a
similar role as the sharp $\Theta$-function in the ``local duality"  wave
function \cite{RadMus97} $\Psi _{\pi }^{A,LD}(x,\vec{k}_{\bot }^{2})
\sim \Theta(\vec k_\bot^2 \leq x\bar x s_0)$, with $s_0=8\pi^2 f_\pi^2$ being
the duality interval. Note that numerically $s_0\approx 0.67$
GeV$^2$ is close to the value of the nonlocality parameter $\lambda_q^2$.
As in the case of (\ref{WF_VF}), the pion light-cone wave function (\ref{Psi})
displays non-analytic dependence on $x$ that disappears if a more realistic
quark-pion vertex with the angular dependence is considered.

\section{Discussion and conclusions.}

Recently, in \cite{Kekez}, it was claimed that the Schwinger-Dyson approach
predicts the same asymptotic coefficient $J\left( \omega \right) =4/3$
for {\it all} nonlocal
quark-photon vertices $\Gamma ^{\mu }\left[ q(k)q(k^{\prime })\gamma (q)
\right] $ which go into the bare ones, $\gamma ^{\mu }$, as soon as one of
the squared momenta ($k^{2}$ or $k^{\prime 2}$) becomes infinite (as in the
Curtis-Pennington~\cite{CPvertex} form of the vertex). In \cite{Kekez}, the
quark propagator that depends on the photon momenta was approximated, at
large $Q^{2}$, by its asymptotic form $\left[ M_{q}^{2}-
\left(k-(q_{1}-q_{2})/2\right)^{2}\right] ^{-1}\rightarrow 2/Q^{2}$. After
this change, the integral (\ref{Ipigg}) attains the same form as the
integral in (\ref{f_pi}) defining $f_{\pi }$. By taking into account the
coefficients in front of the integrals of equations (\ref{f_pi}) and
(\ref{gpiqq}), one immediately reaches the conclusion of \cite{Kekez} (see
also \cite{Dubrovn98}) about the asymptotic coefficient ($J=4/3$). As we
show above, such a quick asymptotic estimation is rather naive and does not
lead to the accurate result. The approximation made in \cite{Kekez} is
justified only in the formal limit $M_{q}>>\Lambda$.

Our analysis is based on the consideration of a triangle diagram in which
the quark propagator and quark-pion vertex are determined nonperturbatively.
In this respect, our approach is close to earlier work \cite{Anselm97}.
However, in \cite{Anselm97}, the approximations in the calculation of the
triangle diagram were used that simplify the dynamics of the process. It
turns out that these approximations are not justified in the kinematic
region of large $|\omega|$. As a result the expression was obtained for
the asymptotic coefficient that is independent of
the internal nonlocal structure of the pion.

In conclusion, within the covariant nonlocal model describing the quark-pion
dynamics, we obtain the $\pi \gamma^{\ast} \gamma^{\ast} $ transition form
factor at moderately high momentum transfers squared, where the perturbative
QCD evolution does not yet reach the asymptotic regime.
 From the model calculations it is possible to find the absolute
normalization of the
asymptotic $Q^{-2}$ term. The asymptotic normalization
coefficient $J( \omega)$ , given in (\ref{Igpigg2}), depends on the ratio of
 the constituent quark mass on a certain soft scale to the
characteristic size of
QCD vacuum fluctuations and also on the kinematics of the process. This
result does not confirm the statement about the universality of the
 asymptotic coefficient given in \cite{Kekez,Dubrovn98,Anselm97}. When
considering
the dependence of the asymptotic coefficient on the internal dynamics, the
CLEO data are consistent with a small value of the diluteness parameter,
which confirms the hypothesis about the small density of the instanton
liquid vacuum~\cite{Shuryak96}. From the comparison of the kinematic
dependence of the asymptotic coefficient of the transition pion form factor,
given by pQCD and the nonperturbative model, the new relation
eq. (\ref{WF_VF}) between the pion distribution amplitude and the dynamical
 quark-pion vertex function is derived. In the specific case of
symmetric kinematics,
our result agrees with the one obtained by OPE~\cite{NSVVZ84}.
The present results are in accordance with the
 conclusions made in \cite{MikhRad90,RadRus96,MikhBak98} within the QCD
sum rules.
A more complete analysis of the light pseudo-scalar meson transition form
factors will be done later, where effects of the finite hadron masses,
nonlocality of the quark-photon vertex, etc., will be considered.

{\centerline{\bf Acknowledgments} }

We are thankful to N.I. Kochelev for discussions on the
 subject of the paper. Our special thanks are to S.V. Mikhailov for
stimulating and helpful discussions. The referee`s remarks were highly
useful. One of us (A.E.D.) thanks the colleagues from
Instituto de F\'{i}sica Te\'{o}rica, UNESP, (S\~{a}o Paulo) for their
hospitality and interest in the work. A.E.D. was partially supported by St.
- Petersburg center for fundamental research grant: 97-0-6.2-28. L.T. thanks
partial support from Conselho Nacional de Desenvolvimento Cient\'{i}fico e
Tecnol\'{o}gico do Brasil (CNPq) and, in particular, to the
 ``Funda\c{c}\~{a} o de Amparo \`{a} Pesquisa do Estado de S\~{a}o Paulo
(FAPESP)'' to provide the essential support for this collaboration.

\begin{figure}[tbp]
\setlength{\epsfxsize}{0.7\hsize} \centerline{\epsfbox{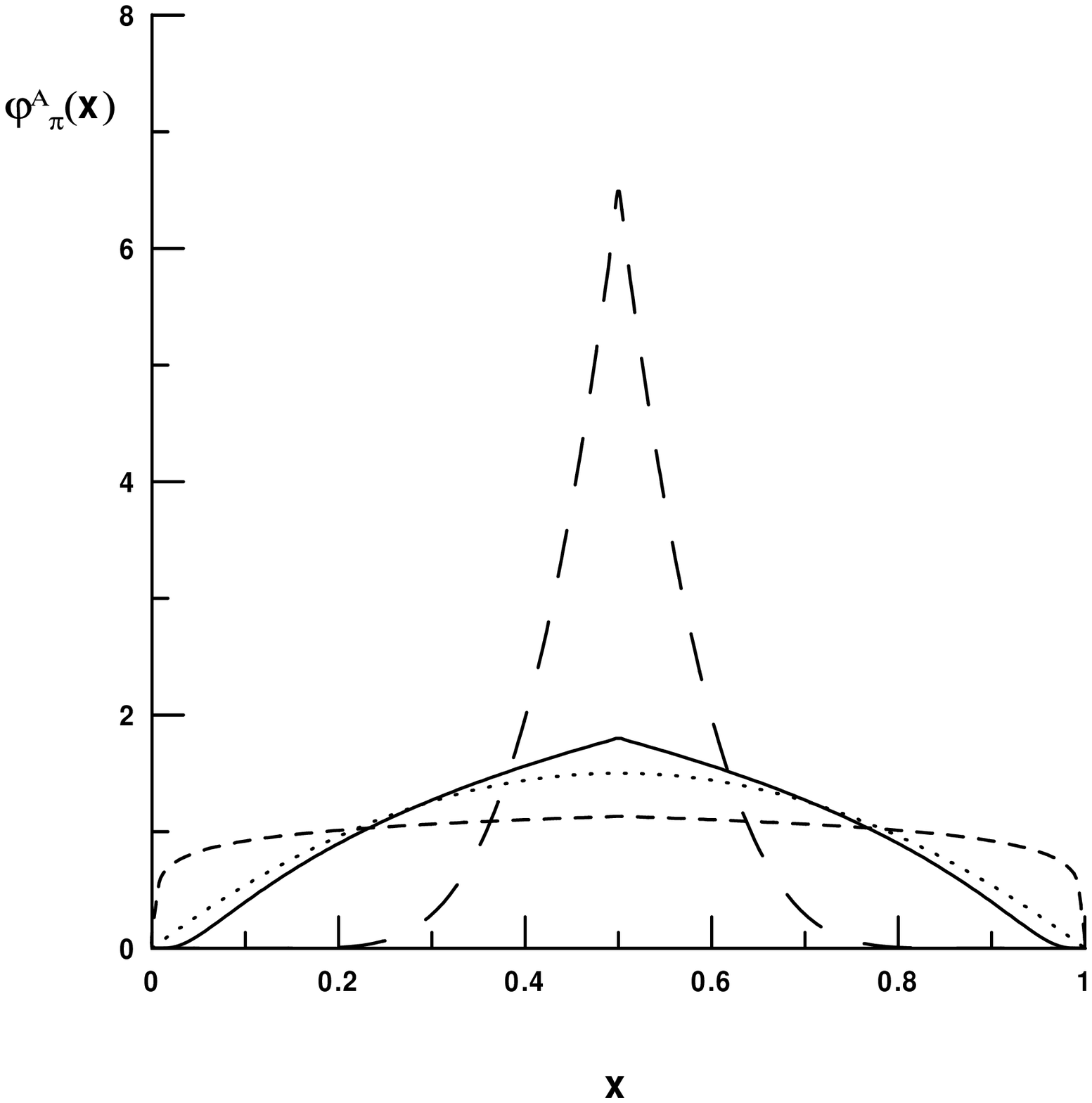}}
\caption[dummy0]{ The pion distribution amplitude as a function of fraction
variable $x$ as given by (\ref{WF_VF}) at different values of the diluteness
parameter: $\protect\chi^2=0.15$ (solid line), $\protect\chi^2=0.0001$
(short-dashed line), $\protect\chi^2=4$ (long-dashed line). The asymptotic
distribution amplitude is marked by point line.}
\end{figure}

\begin{figure}[tbp]
\setlength{\epsfxsize}{0.8\hsize} \centerline{\epsfbox{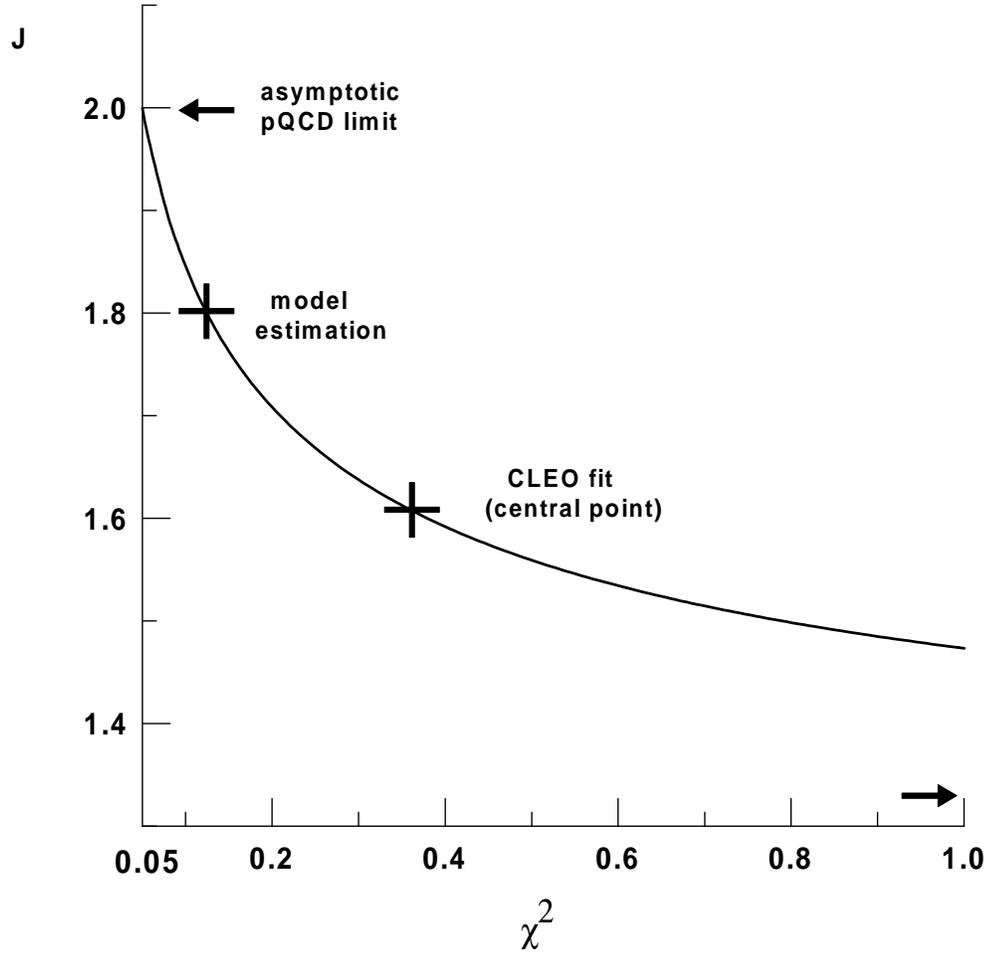}}
\caption[dummy0]{ The asymptotic coefficient $J_{np}$ as function of the
diluteness parameter squared $\protect\chi^2$, for the process $\protect
\gamma\protect\gamma^\ast\to\protect\pi^0$ ($\protect\omega=1$). It is
indicated the values of $J_{np}$ obtained from the fit of CLEO data (central
 point), and the predictions obtained from nonperturbative covariant
model ($ \protect\chi^2=0.15$), pQCD~\protect\cite{BrLep79}; the right
arrow points to the limiting value of $J=4/3$ at $\protect\chi^2\to
\infty$.}
\end{figure}

\begin{figure}[tbp]
\setlength{\epsfxsize}{0.8\hsize} \centerline{\epsfbox{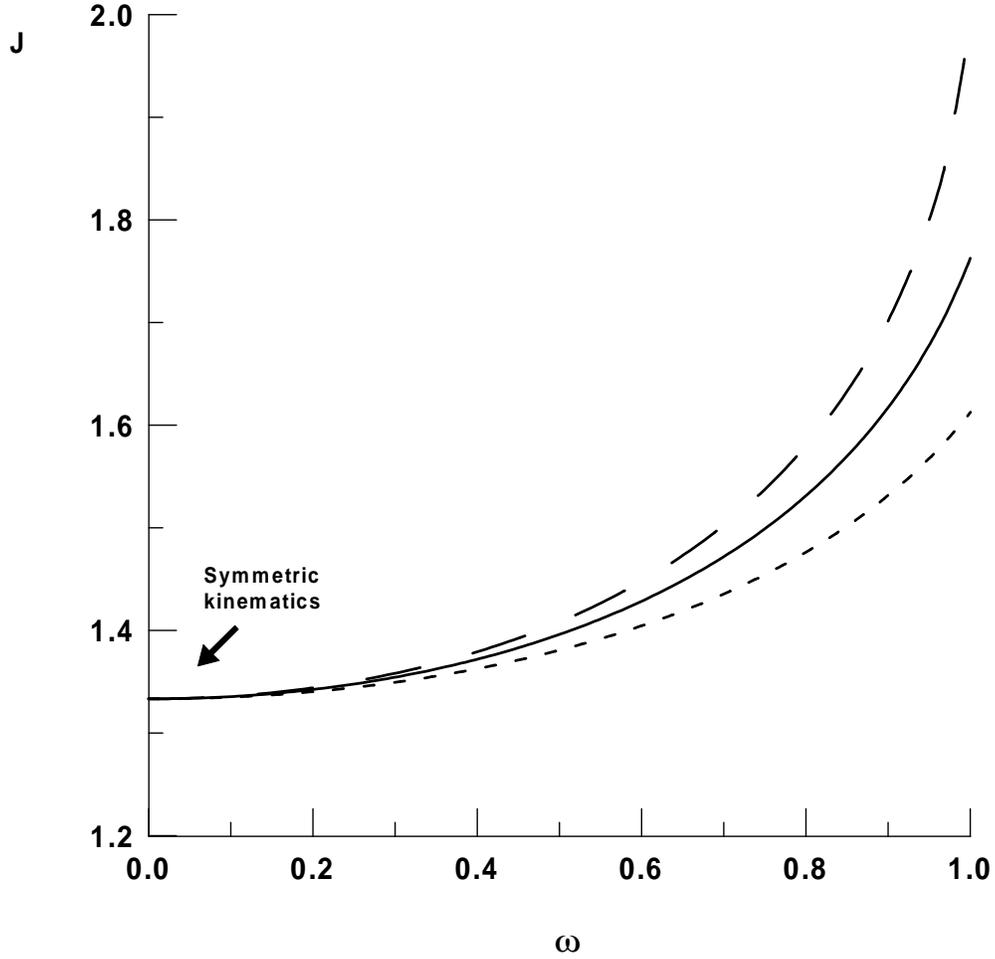}}
\caption[dummy0]{The asymptotic coefficient $J_{np}$ as a function of the
kinematic parameter $\protect\omega$. Solid line corresponds to
$\protect \chi^2=0.15$, giving $J_{np}(\omega=1)=1.8$. Short-dashed line
 is for $\protect\chi^2=0.35$, giving $J_{np}(\omega=1)=1.6$.
Long-dashed line corresponds to asymptotic pQCD prediction given by
(\ref{J}) with $J_{asympt}(\omega=1)=2$. It is
also indicated $J_{np}$ for symmetric kinematics $q_1^2 = q_2^2$.}
\end{figure}

\end{document}